\documentclass[useAMS,usenatbib,letters]{mn2e}
\usepackage{aas_macros,graphicx,times,multirow}
\title[Spin-down rate of \src]
  {Spin-down rate and inferred dipole magnetic field of the soft gamma-ray repeater SGR\,1627--41}
\author[P. Esposito et al.]
{P.~Esposito,$^{1,2}$\thanks{E-mail: paoloesp@iasf-milano.inaf.it} M.~Burgay,$^{3}$  A.~Possenti,$^{3}$ R.~Turolla,$^{4,5}$ S.~Zane,$^{5}$ 
\newauthor A.~De~Luca,$^{1,6}$ A.~Tiengo,$^{1}$ G.~L.~Israel,$^{7}$ F.~Mattana,$^{8}$ S.~Mereghetti,$^{1}$
\newauthor M.~Bailes,$^{9}$ P.~Romano,$^{10}$ D.~G\"otz,$^{11}$ and N.~Rea$^{12,13}$\\
$^1$INAF/Istituto di Astrofisica Spaziale e Fisica Cosmica - Milano, via E.~Bassini 15, 20133 Milano, Italy\\
$^2$Istituto Nazionale di Fisica Nucleare, sezione di Pavia, via A.~Bassi 6, 27100 Pavia, Italy\\
$^3$INAF/Osservatorio Astronomico di Cagliari, localit\`a Poggio dei Pini, strada 54, 09012 Capoterra, Italy\\
$^4$Universit\`a degli Studi di Padova, Dipartimento di Fisica, via F.~Marzolo 8, 35131 Padova, Italy\\
$^5$Mullard Space Science Laboratory, University College London, Holmbury St. Mary, Dorking, Surrey RH5 6NT, UK\\
$^6$IUSS - Istituto Universitario di Studi Superiori, viale Lungo Ticino Sforza 56, 27100 Pavia, Italy\\
$^7$INAF/Osservatorio Astronomico di Roma, via Frascati 33, 00040 Monteporzio Catone, Italy\\
$^{8}$Laboratoire AstroParticule et Cosmologie, Universit\'e Paris 7 - Denis Diderot, rue A.~Domon et L.~Duquet 10, 75205 Paris, France\\
$^{9}$Centre for Astrophysics and Supercomputing, Swinburne University of Technology, PO Box 218, Hawthorn, VIC 3122, Australia\\
$^{10}$INAF/Istituto di Astrofisica Spaziale e Fisica Cosmica - Palermo, via U.~La Malfa 153, 90146 Palermo, Italy\\
$^{11}$CEA Saclay, DSM/Irfu/Service d'Astrophysique, Orme des Merisiers, B\^at. 709, 91191 Gif sur Yvette, France\\
$^{12}$University of Amsterdam, Astronomical Institute Anton Pannekoek, Kruislaan 403, 1098~SJ Amsterdam, The Netherlands\\
$^{13}$CSIC - Institut d'Estudis Espacials de Catalunya, Campus UAB, Fac. Ci\`encies, Torre C5 parell 2. 08193 Bellaterra, Spain}
\date{Accepted 2009 July 14. Received 2009 July 14; in original form 2009 July 3}
\pagerange{\pageref{firstpage}--\pageref{lastpage}} \pubyear{2009}
\def\LaTeX{L\kern-.36em\raise.3ex\hbox{a}\kern-.15em
    T\kern-.1667em\lower.7ex\hbox{E}\kern-.125emX}
\def\xmm {\emph{XMM-Newton}}
\def\cxo {\emph{Chandra}}
\def\swift {\emph{Swift}}
\def\sax {\emph{BeppoSAX}}

\def\asca {\emph{ASCA}}
\def\src {SGR\,1627--41}
\def\flux {\mbox{erg cm$^{-2}$ s$^{-1}$}}
\def\lum {\mbox{erg s$^{-1}$}}

\begin{document}
\label{firstpage}
\maketitle
\begin{abstract}
Using \cxo\ data taken on 2008 June, we detected pulsations at 2.594\,39(4) s in the soft gamma-ray repeater \src. This is the second measurement of the source spin period and allows us to derive for the first time a long-term spin-down rate of $(1.9 \pm 0.4)\times10^{-11}$ s s$^{-1}$.  From this value we infer for \src\ a characteristic age of $\sim$2.2 kyr, a spin-down luminosity of $\sim$$4\times10^{34}$ \lum\ (one of the highest among sources of the same class), and a surface dipole magnetic field strength of $\sim$$2\times10^{14}$ G. These properties confirm the magnetar nature of \src; however, they should be considered with caution since they were derived on the basis of a period derivative measurement made using two epochs only and magnetar spin-down rates are generally highly variable. The pulse profile, double-peaked and with a pulsed fraction of ($13\pm 2$)\% in the 2--10 keV range, closely resembles that observed by \xmm\ in 2008 September. Having for the first time a timing model for this SGR, we also searched for a pulsed signal in archival radio data collected with the Parkes radio telescope nine months after the previous X-ray outburst. No evidence for radio pulsations was found, down to a luminosity level $\sim$10--20 times fainter (for a 10\% duty cycle and a distance of 11 kpc) than the peak luminosity shown by the known radio magnetars.
\end{abstract}

\section{Introduction}
Anomalous X-ray pulsars (AXPs) and soft gamma-ray repeaters (SGRs) are isolated neutron stars with periods of several seconds ($P\sim2$--12 s), rapid spin down ($\dot{P}\sim10^{-11}$ s s$^{-1}$), bright ($\sim$$10^{34}$--$10^{35}$ \lum) and highly variable X-ray emission.\footnote{Ten AXPs and six SGRs are confirmed, and there are a few candidates; see catalog at \mbox{http://www.physics.mcgill.ca/$\sim$pulsar/magnetar/main.html}.} 
AXPs and SGRs are commonly interpreted in terms of the \emph{magnetar} model. Magnetars are ultra-magnetized neutron stars with magnetic fields largely in excess of the quantum critical field $B_{\rm{QED}}=\frac{m_e^2c^3}{\hbar e}\simeq 4.4$$\times$$10^{13}$ G \citep{paczynski92,duncan92,thompson95,thompson96}. Contrary to what happens in ordinary radio pulsars, the X-ray luminosity is larger than their rotational energy loss. Since no stellar companions have been detected thus far, also accretion is unlikely to be responsible for the emission of AXPs and SGRs. Their persistent X-ray luminosity, as well as the bursts and flares typical of these sources, are instead believed to be powered by the decay of their ultra-strong magnetic field (see \citealt{woods06} and \citealt{mereghetti08} for recent reviews).\\
\indent AXPs were first recognised as a class of persistent X-ray pulsars, with the peculiarity that the X-ray luminosity exceeds that available from spin-down (whence the name ``anomalous''; \citealt{mereghetti95}). SGRs were first noticed as hard-X/$\gamma$-ray transients \citep{laros87}, characterized by recurrent, short ($<$1 s) and relatively soft (peak photon energy $\sim$25--30 keV) flashes with super-Eddington luminosity. Although SGRs and AXPs have been discovered through very different channels, observations performed over the last few years highlighted several similarities among these two classes of objects and pointed towards a common magnetar nature (see e.g. \citealt{rea09}). In particular, short and hard X-ray bursts, originally considered as the defining characteristic of SGRs, have now been observed in several AXPs (see e.g. \citealt*{gavriil02}; \citealt{mereghetti09}).\\
\indent \src\ was discovered in 1998, when about one hundred bursts in six weeks were observed by \emph{CGRO}/BATSE and other instruments \citep{woods99}. Its soft X-ray counterpart was identified with \sax\ in 1998 at a luminosity level of $\sim$$10^{35}$ \lum\ \citep{woods99}.\footnote{Here and through the paper we assume a distance to the source of 11 kpc ($d=11.0\pm0.3$ kpc; \citealt{corbel99}).} Subsequent observations carried out with \sax, \asca, \cxo, and \xmm\ showed a spectral softening and a monotonic decrease in the luminosity, down to a level of $\sim$$10^{33}$ \lum\ \citep{kouveliotou03,mereghetti06,eiz08}.\\
\indent After nearly ten years of quiescence, \src\ re-activated on 2008 May 14, when several bursts were detected by \swift/BAT and other hard X-ray instruments \citep{eiz08}. The burst re-activation was associated with a large enhancement of the soft X-ray flux and a marked spectral hardening.\\
\indent Until very recently, \src\ was the only magnetar candidate with no pulsation period known. In order to search in depth for pulsations taking advantage of the high flux state, we asked for a long \xmm\ observation to be carried out during its outburst. The observation was performed on 2008 September 27--28 and we could detect a clear pulsation period of 2.594\,578(6) s \citep{esposito09}. However, no meaningful constraints on the period derivative could be derived from that observation.\\
\indent Here we report on a new measurement of the period using data gathered shortly after the burst activation by the \emph{Chandra X-ray Observatory}. This allows us to estimate for the first time the spin-down rate of \src\ and to infer its magnetic field, characteristic age, and spin-down luminosity. Taking advantage of the new pieces of information about the timing properties of \src, we also searched for a pulsed signal in archival radio data collected at the Parkes observatory.

\section{The \emph{Chandra} observation: data analysis and results}
The \emph{Chandra X-ray Observatory} \citep{weisskopf00} pointed its mirror towards \src\ on 2008 June 3 (MJD 54620) and observed the source for about 40 ks (observation identifier: 9126). The observation was carried out with the Advanced CCD Imaging Spectrometer (ACIS; \citealt{garmire03}) instrument operated in the Continuous Clocking (CC) mode, which provides a time resolution of 2.85 ms and imaging along a single direction. The event telemetry was in Faint mode. The source was positioned in the back-illuminated ACIS-S3 chip, sensitive to photons in the 0.2--10 keV energy range.\\
\indent The data were processed using the \cxo\ Interactive Analysis of Observation software (\textsc{ciao}, version 4.1) and we employed the most updated calibration files available at the time the reduction was performed (\textsc{caldb} 4.1). Standard screening criteria were applied in the extraction of scientific products.\footnote{See the \cxo\ Science Threads at the \cxo\ X-ray Center web site, \mbox{http://asc.harvard.edu/ciao/threads/index.html}.} No significant background flares affected the observation.\\
\indent The source photons for the timing and spectral analyses were accumulated from a $5\times5$ pixels region centred on \src\ (one ACIS-S pixel corresponds to $0\farcs492$); the background events were extracted from source-free regions of the same chip as the source. A total of about $1120\pm40$ counts above the background were collected from \src\ in the 2--10 keV energy range. 
\subsection{Spectroscopy}\label{spectroscopy}
The ancillary response file and the redistribution matrix for the spectral fitting were generated with the \textsc{ciao} tasks \textsc{asphist}, \textsc{mkarf}, and \textsc{mkacisrmf}, using the specific bad-pixel file of this observation. The data were grouped with a minimum of 20 counts per energy bin and the spectrum was analysed with the \textsc{xspec} version 12.4 analysis package \citep{arnaud96}.\\
\indent Given the paucity of counts, we fit a simple model to the data: a power law corrected for interstellar absorption. We obtained the following best-fit parameters ($\chi^2_{\rm{r}}=1.13$ for 51 degrees of freedom): absorption $N_{\rm H}=10^{+1}_{-2}\times10^{22}$ cm$^{-2}$ and photon index \mbox{$\Gamma=1.0^{+0.3}_{-0.2}$} (here and in the following all errors are at 1$\sigma$ confidence level). The absorbed 2--10 keV flux was $\sim$$1.3\times10^{-12}$ \flux, corresponding to a luminosity of $\sim$$3\times10^{34}$ \lum. These results are consistent with those reported in \citet{woods08atel1564} and confirm the bright and hard state of the source following the 2008 May 28 burst activation \citep{eiz08,esposito09}.
\subsection{Timing}
For the timing analysis, the photon arrival times were converted to the Solar System barycentre with the \textsc{ciao} task \textsc{axbary} using the source coordinates reported in \citet{wachter04}.
We searched for the presence of a periodic signal using a $Z_2^2$ test (see \citealt{esposito09}) over the period range $2.584\,71$--$2.594\,60$ s; this range was determined by extrapolating from the 3$\sigma$ lower limit on the value reported in \citet{esposito09}, conservatively assuming a period derivative of $0\leq\dot{P}\leq10^{-9}$ s s$^{-1}$. The period search step size was $\sim$$8\times10^{-6}$ s, which is equivalent to oversampling the Fourier period resolution ($\frac{1}{2}P^2/T_{\rm{obs}}$) by a factor of 10.
\begin{figure}
\resizebox{\hsize}{!}{\includegraphics[angle=0]{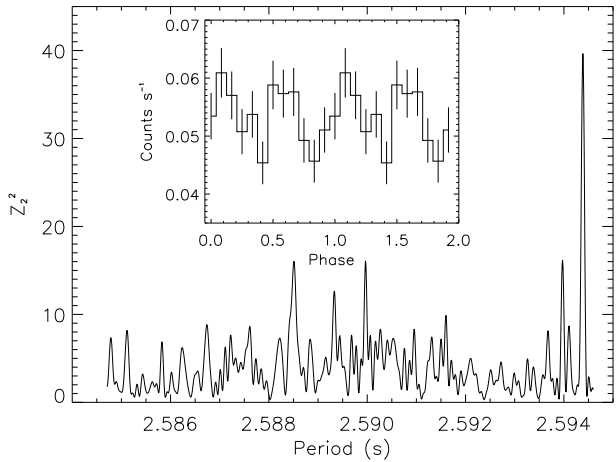}}
\caption{\label{timing}  Z$^2_2$-periodogram of \src\  in the 2--10 keV energy range. The peak at 2.594\,39 s is significant at 4$\sigma$. The inset shows the corresponding pulse profile (2--10 keV, not background-subtracted).}
\end{figure}\\
\indent A significant signal was found in the $Z_2^2$-periodogram at $\sim$2.594\,39 s (see Figure \ref{timing}). The probability of this peak (with a $Z_2^2$ value of 39.66) to appear by chance in the search, taking into account the number of trials (1177), is $6\times10^{-5}$. This corresponds to a 4-sigma detection. To refine our period estimate, we used an epoch folding technique and fitted the peak in the $\chi^2$ versus trial period distribution as described in \citet{leahy87}. We obtained a best period of $2.594\,39 \pm 0.000\,04$ s. The corresponding folded lightcurve is shown in Figure \ref{timing}; the pulse profile is double-peaked and the root mean square pulsed fraction is ($13 \pm 2$)\% in the 2--10 keV energy range and after subtracting the background. The period derivative inferred from the \cxo\ and \xmm\ measurements is $(1.9 \pm 0.4)\times10^{-11}$ s s$^{-1}$. Assuming that the spin-down rate has remained constant at this value, we repeated the search for pulsations in archival X-ray data described in \citet{esposito09}. Again, we did not detect any significant signal.\\
\indent To search for possible pulse shape variations as a function of time, we compared the \cxo\ lightcurve with that obtained in 2008 September with \xmm\ by using a two-dimensional Kolmogorov--Smirnov test \citep{peacock83,fasano87}. Taking into account the unknown relative phase alignement,\footnote{The accuracy of the timing solution does not allow us to phase-connect the \xmm\ and \cxo\ data.} the two profiles are compatibile. In fact, the probability that they come from the same underlying distribution is about 70\%.

\section{Search for radio pulsations}
\begin{figure}
\resizebox{\hsize}{!}{\includegraphics[angle=0]{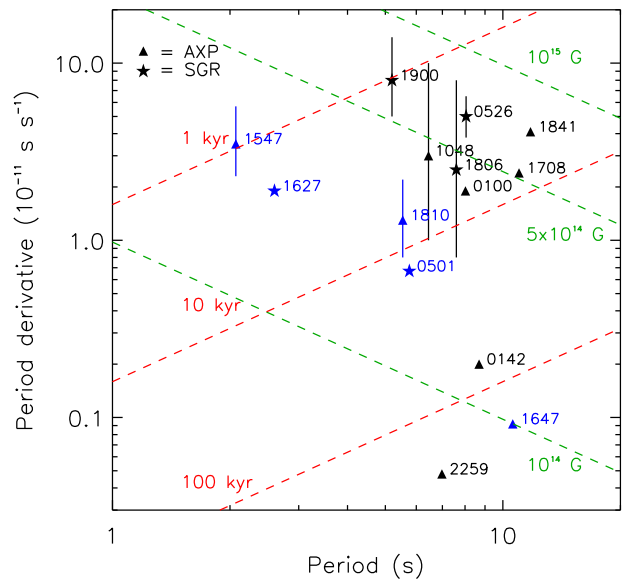}}
\caption{\label{ppdot} $P$--$\dot{P}$ diagram for the known magnetars; vertical bars indicate variability ranges. Lines of constant magnetic field strength (green) and characteristic age (red) are indicated. The `transient' magnetars, the ones that show the largest variations of the persistent X-ray flux (up to two or three orders of magnitude), are plotted in blue. Data are from \citet{mereghetti08} and our analyses.}
\end{figure}

The $P$--$\dot{P}$ diagram for magnetars (Figure \ref{ppdot}) shows that the timing properties of \src\ are remarkably similar to those of the anomalous X-ray pulsar 1E\,1547.0--5408.
The latter source, together with XTE\,J1810--197 \citep{camilo06}, is one of the two magnetars known to sporadically emit radio pulses \citep{camilo07,camilo08,burgay09}. Although \src\ was not detected as a radio pulsar immediately after the 2008 May activation \citep{camilo08atel1558}, we searched for radio emission in archival data taking advantage of the new pieces of information about its timing properties.\\
\indent We analysed archival radio observations performed at 1.4 GHz with the Parkes radio telescope. The data were taken on 1999 March 22 with the central beam of the 20-cm multibeam receiver \citep{staveley96} over a bandwidth of 288 MHz split in 96 3-MHz channels. The 2.3 hours observation was 1-bit sampled every 1 ms.\\
\indent We folded the data with 6250 values of the period spanning $\pm$5 ms (corresponding to a $\sim 4 \sigma$ uncertainty on the value of $P$) around the nominal value $P_{\rm{PKS}}=2.5889(12)$ s extrapolated from the current best X-ray ephemeris at MJD 51259, which corresponds to the epoch of the Parkes observation. Given the position in the sky of \src, assuming a distance of 11 kpc \citep{corbel99} and a model for the distribution of free electrons in the interstellar medium \citep{cordes02}, the expected dispersion measure is DM $\sim1150$ pc cm$^{-3}$. Given the uncertainties in the DM determination, we chose to de-disperse the signal with 390 DM values ranging from 0 to 2300 pc cm$^{-3}$. The expected broadening of the pulse due to interstellar scattering at \src\ position, according to the Cordes \& Lazio model \citep{cordes02}, is $\sim$50 ms at 1.4 GHz; the uncertainties of the interstellar medium model in this respect are, however, even larger than those related to the DM. The number of period and DM steps was hence chosen in such a way to produce a maximum total smearing in the folded profile of $< $10 ms, also compatible with the number of bins $n_{\rm{bin}}=256$ in which the folded profile was subdivided. No signal with signal-to-noise ratio greater than 6 was found in this search. Using the radiometer equation (e.g. \citealt{manchester01}) we find an upper limit for radio pulsed emission of 0.22 mJy for an approximately sinusoidal pulse profile, and of 0.08 mJy for a duty cycle of 10\%.\\
\indent Since the two known radio-pulsating magnetars are sometimes visible through their individual pulses \citep{camilo07}, also a search for single dispersed pulses has been carried out, leading to the detection of a faint (signal-to-noise ratio: 6.2) candidate signal at DM $\sim93$ pc cm$^{-3}$. The DM of the putative pulse (to be confirmed with further observations) is however likely too small to be associated with \src. A blind search for periodic signals at DM = 93 pc cm$^{-3}$ resulted in no significant detection down to a flux density limit of $\sim$0.01 mJy for a long period pulsar and $\sim$0.2 mJy for a millisecond pulsar.

\section{Discussion and conclusions}\label{disc}
With a rotation period of 2.59 s \citep{esposito09}, \src\ is the second fastest spinning magnetar, after 1E\,1547.0--5408 ($P=2.07$ s; \citealt{camilo07}). Using an archival \cxo\ observation, we have been able to obtain a second period measurement. The \cxo\ and \xmm\ datasets, separated by about 114 days, imply a long-term average spin-down rate $\dot{P} = (1.9 \pm 0.4)\times10^{-11}$ s s$^{-1}$. This value is compatible with the range $1.2\times10^{-11}$ s s$^{-1}<\dot{P}<6\times10^{-10}$ s s$^{-1}$ derived from the long \xmm\ observation.\\
\indent Within the usual vacuum dipole framework (see e.g. \citealt{lorimer04}), the spin-down rate can be used to infer a surface magnetic field strength of $B\approx3.2\times10^{19}(P\dot{P})^{1/2}\simeq2\times10^{14}$ G, confirming the magnetar nature of \src. 
The characteristic age and the spin-down luminosity are \mbox{$\tau_c=\frac{1}{2}P/\dot{P}\simeq2.2$ kyr} and $\dot{E}=4\pi^2 I \dot{P}P^{-3}\simeq4\times10^{34}$ \lum, respectively, where $I\approx10^{45}$ g cm$^2$ is the moment of inertia of the neutron star. \\
 \indent Our newly determined values of $P$ and $\dot P$ for \src\ are reported in  Figure \ref{ppdot} together with all the values available up to now for  magnetar sources. As can be seen in Figure \ref{ppdot}, where the vertical bars indicate variability ranges, magnetar spin-down rates can be highly variable. For this reason the magnetic fields, characteristic ages, and spind-down luminosities inferred for magnetars should be taken with particular caution. In particular, this applies to \src, for which we have no information of possible variability of the period derivative. It is interesting to note that SGRs and AXPs do not populate different regions of the $P$--$\dot P$ plane, so that it would be difficult to discriminate between the two groups on the basis of their timing properties. This further supports the idea that SGRs and AXPs are actually members of the same class. What Figure \ref{ppdot} suggests, instead, is that {\em transient} and {\em persistent} sources might have different characteristics. The transient magnetars (in blue in Figure \ref{ppdot}), in fact, appear to have lower  magnetic fields with respect to the persistent ones (irrespectively of their classification, AXPs or SGRs). In this respect the position of \src\ in the $P$--$\dot P$ diagram is similar to that of other transient magnetar sources. The only exceptions are 4U\,0142+614 and 1E\,2259+586 which are not transient but have among the lowest derived values of $B$.\\
\indent The neutron-star characteristic age is consistent with an association of \src\ with the supernova remnant (SNR) G337.0--0.1 (see \citealt{esposito09} and references therein). At a distance of 11 kpc \citep{corbel99}, the observed SNR angular diameter of $\sim$3 arcmin would correspond to a physical diameter of $\sim$9--10 pc. This is similar to the observed sizes for other young remnants ($\sim$1--3 kyr, at the beginning of the Sedov phase) hosting a neutron star, like for example Kes\,73 \citep{gotthelf97} and RCW\,103 \citep{carter97}. \\
\indent The spin-down luminosity of \src\ is one of the highest among magnetars and is roughly equal to the \cxo\ luminosity of $3\times10^{34}$ \lum\  (\S\,\ref{spectroscopy}). In magnetars $\dot{E}$ ranges, in fact, from $6\times10^{31}$ \lum\ for 1E\,2259+586 \citep{gk02} to $10^{35}$ \lum\ for 1E\,1547.0--5408 \citep{camilo07}. The observed correlation between the spin-down power and the nonthermal X-ray emission for ordinary pulsars by \citet{possenti02} predicts for \src\ an X-ray luminosity of $\approx$$10^{31}$ \lum\ and a maximum value $L_{\rm{X,crit}}=10^{-18.5}\big(\frac{\dot{E}}{[\rm{erg\ s^{-1}}]}\big)^{1.48}$ \lum\ $\simeq5\times10^{32}$ \lum. This makes it implausible that \src\ is powered by star rotation unless the conversion efficiency is extremely high.
Moreover, \citet{camilo07} noted in the case of 1E\,1547.0--5408 that, despite the fact that the spin-down luminosity is comparable with the X-ray luminosity, it is unlikely that a significant fraction of the X-ray emission is powered by rotation, since the source displays the distinctive features of the pulsars powered by magnetic field decay. In fact, at variance with rotation-powered pulsars\footnote{With the notable exception of the young pulsar PSR\,J1846--0258 at the centre of the supernova remnant Kes\,75, which showed large flux and spectral variations \citep{gavriil08}. This source, however, shares some (but not all) characteristics of magnetars, including a (relatively) high dipole magnetic field of $5\times10^{13}$ G and the emission of SGR-like short bursts \citep{gavriil08}, and it is therefore unclear whether it is purely rotation-powered.} and cooling neutron stars, the source showed flux variations by orders of magnitude. Furthermore, the X-ray spectrum of 1E\,1547.0--5408 includes a thermal component which is hotter ($kT\simeq0.4$ keV) than that expected from a young, cooling neutron star (see for example \citealt{yakovlev02}). Such high temperatures are instead typical of magnetars, the surface of which can be substantially heated by energy deposition following burst-active periods and/or by dissipative currents in the crust. The considerations by \citet{camilo07} apply also to \src, with its X-ray luminosity variable by a factor of $\sim$100 (between the historical minimum and maximum, $\sim$$2\times10^{33}$ \lum\ and $\sim$$3\times10^{35}$ \lum; \citealt{eiz08}); also, the only high-statistics spectrum of \src, collected by \xmm\ in 2008 September, when the luminosity was still high ($\sim$$10^{34}$ \lum), suggests the presence of a thermal component, a blackbody with temperature $kT\simeq0.5$ keV \citep{esposito09}. Finally, we also note that it is unlikely that the low luminosity observed in \src\ while it was approaching its historical minimum during its decade-long quiescence arose mainly from rotational power. The X-ray emission was in fact very soft and possibly thermal \citep{kouveliotou03,mereghetti06}.\\
\indent The mechanisms behind the pulsed radio emission of magnetars are poorly understood. It is not ruled out that their radio emission is related to the braking in a way similar to that of the ordinary radio pulsars, but the radio properties of the two magnetars detected so far in radio, XTE\,J1810--197 and 1E\,1547.0--5408, are quite distinguishing: their flux is highly variable on daily timescales, their spectrum is very flat, and their average pulse profile changes with time, from minutes to days \citep{camilo06,camilo07,camilo08,kramer07}. Remarkably, XTE\,J1810--197 and 1E\,1547.0--5408 share small periods and high spin-down luminosities. Thus, in view of the similarities with the two radio-pulsating magnetars, \src\ is a good candidate to be searched for pulses at radio frequencies.\\
\indent Our analysis of archival Parkes data obtained at 1.4 GHz in 1999 March (about 9 months after the first detected X-ray outburst of \src) showed no pulsed signal in a $\pm$4$\sigma$ interval bracketing the expected period, down to a flux density limit of $S\simeq 0.08$ mJy for a pulsar with a 10\% duty cycle (twice as better a limit than what obtained by \citet{camilo08atel1558} with observations following the 2008 outburst). For a distance of $d=11$ kpc \citep{corbel99}, this limit translates into a pseudo-luminosity $L=Sd^2$ of approximately 10 mJy kpc$^2$, significantly smaller than the 1.4 GHz luminosity of the two known radio magnetars at their peak ($\sim$100--200 mJy kpc$^2$), although still much larger than the smallest known luminosity of ordinary young (with spin-down age $\tau_c < 10^5$ yr) pulsars ($\sim$1 mJy kpc$^2$). Given the remarkable and rapid variability of the pulsed flux shown by the two aforementioned known radio magnetars, the negative results of the two searches performed so far on \src\ (\citealt{camilo08atel1558} and this work) cannot anyway be conclusive and a longer term monitoring is necessary for satisfactorily assessing the radio properties of this source.

\section*{Acknowledgments}
This research has made use of data obtained from the \cxo\ Data Archive and software provided by the \cxo\ X-ray Center (CXC) in the application package \textsc{ciao}. The Parkes radio telescope is part of the Australia Telescope which is funded by the Commonwealth of Australia for operation as a National Facility managed by CSIRO. The authors thank the referee, Vicky Kaspi, for her constructive comments. The Italian authors acknowledge the partial support from ASI (ASI/INAF contracts I/088/06/0 and AAE~TH-058). SZ acknowledges support from STFC. FM and DG acknowledge the CNES for financial support. NR is supported by an NWO Veni Fellowship. 

\bibliographystyle{mn2e}
\bibliography{biblio}

\bsp

\label{lastpage}

\end{document}